\documentclass[%
 reprint,
 amsmath,amssymb,
 aps,
physrev
floatfix,
]{revtex4-2}

\usepackage{graphicx}
\usepackage{dcolumn}
\usepackage{bm}
\usepackage{hyperref}

\def\Tr{\mathrm{Tr}}

\def\FE{A}

\begin{document}


\title{Efficient Bosonic and Fermionic Sinkhorn Algorithms for Non-Interacting Ensembles in One-body Reduced Density Matrix Functional Theory in the Canonical Ensemble}

\author{Derk P. Kooi}%
 \email{derkkooi@gmail.com}
\affiliation{%
 Department of Chemistry and Pharmaceutical Sciences and Amsterdam Institute for Molecular and Life Sciences (AIMMS), Faculty of Science, Vrije Universiteit, De Boelelaan 1083, 1081HV Amsterdam, The Netherlands
}%

\date{\today}

\begin{abstract}
We introduce 1-RDMFT in the canonical ensemble and then proceed to approximate the interacting ensemble by a non-interacting ensemble that maximizes the entropy, independently of temperature. Bosonic and Fermionic Sinkhorn algorithms are derived and used to invert the relationship between the Natural Orbital Occupation Numbers (NOONs) and the effective orbital energies of the non-interacting ensemble. Both the Bosonic and Fermionic Sinkhorn algorithms are shown to perform well in reproducing the NOONs of simulated distributions and the ground-state NOONs of H$_2$O and H$_2$. In the case of H$_2$ we also study the resulting non-interacting entropy and non-interacting approximation to the interaction energy within several wavefunction subspaces as the bond length varies. This provides several new starting points for approximations of the interaction energy, also at zero-temperature. Connections to entropically-regularized Multi-Marginal Optimal Transport (MMOT) are highlighted that may prove interesting for future research.
\end{abstract}

                              
\maketitle

\section{Introduction}
Recently, interest has risen in utilizing 1-body Reduced Density Matrix Functional Theory (1-RDMFT) at finite temperature.\cite{BalCanGro-PRA-15,  GieRug-PR-19} This interest does not only arise from the desire of studying physical systems at temperatures in which thermal effects are important, such as Warm dense Matter, but also from the fact that the entropic term acts to regularize the calculation and allows for a simplified computational procedure.\cite{BalGro-CTC-13} The 1-RDM $\gamma$ is defined as,
\begin{equation}
\gamma_{pq} = \mathrm{Tr}( \hat{\Gamma}\hat{a}^\dagger_p \hat{a}_q),
\end{equation}
where $\hat{\Gamma}$ is the density operator of the many-body system and $\hat{a}^\dagger_p$ and $\hat{a}_q$ are bosonic or fermionic creation and annihilation operators. The basis in which $\gamma$ is diagonal is called the Natural Orbital (NO) basis and the eigenvalues of $\gamma_{pq}$ are called the Natural Orbital Occupation Numbers (NOONs). In a system at finite temperature they satisfy $0 < n_p (< 1)$, where the upper-bound only holds for fermions.

Furthermore, the (grand canonical) non-interacting entropy of the 1-RDM at an effective temperature has recently been successfully used to approximate the correlation energy corresponding to the 1-RDM at zero temperature along the dissociation curves of diatomic molecules with only two parameters per molecule (the effective temperature and an inconsequential constant shift).\cite{WanBae-PRL-22} The crucial advantage of this approach to electron correlation is that the cost becomes essentially that of a Hartree-Fock Self-Consistent Field (SCF) procedure, while previous approaches to 1-RDMFT at zero temperature often come at a significantly increased computational cost w.r.t. Hartree-Fock due to issues with the convergence of the calculation.\cite{CanPer-JCP-08}

It remains an open question if the convergence of more elaborate functionals is also improved by including the non-interacting entropy. The inclusion of the non-interacting entropy allows for a direct determination of the NOONs in terms of the NO energies $\{\epsilon_p\}$, removing them as variables to be optimized over, such that only the NO's need to be determined. Recent work has shown that second-order algorithms can drastically reduce the number of iterations needed to converge 1-RDMFT calculations compared to iterative-diagonalization (e.g. SCF) procedures.\cite{ElaGupHol-JCP-22} Combining these approaches may lead to new 1-RDMFT procedures with robust convergence.

The advantage of 1-RDMFT over a description at the level of the full $N$-body wavefunction arises from the fact that the 1-RDM is a one-body object and therefore does not suffer from the curse of dimensionality. If we fix the two-body interaction we can write the total energy as a universal functional of the 1-RDM, where in practice we need to find approximations to the interaction energy as a functional of the 1-RDM.\cite{Gil-PR-75} At finite temperature we define the following free energy ($F^\beta[\gamma]$) and grand potential ($\Omega^{\beta, \mu}[\gamma]$) functionals, where $\beta=\frac{1}{k_B T}$ is the inverse temperature and $\mu$ is the chemical potential,
\begin{align}
\FE^\beta[\gamma] &:= h[\gamma] + \min_{\mathcal{T}\otimes \mathcal{T} \ni \hat{\Gamma} \rightarrow \gamma} \left( \hat{\Gamma} \hat{W} + \frac{1}{\beta} \mathrm{Tr}(\hat{\Gamma} \log(\hat{\Gamma}) \right) \nonumber \\
&=: h[\gamma] + \FE^\beta_\mathrm{int}[\gamma], \label{eq:functional},\\
\Omega^{\beta, \mu}[\gamma] &:= \FE^\beta[\gamma] - \mu N[\gamma],
\end{align}
where $h[\gamma]$ is the energy obtained from a given one-body Hamiltonian $\hat{h}$ (e.g. kinetic energy, interaction with external potential) and $N[\gamma]$ is the particle number. Explicitly $h[\gamma] = \Tr(\gamma h)$ and $N[\gamma] = \Tr(\gamma)$. $\hat{W} = \frac{1}{2}\sum_{pqrs} a^\dagger_p a^\dagger_q a_s a_r \langle pq | rs \rangle$ is the operator corresponding to the two-particle interaction, where we will leave $\langle pq | rs \rangle$ unspecified. We then minimize the free energy (canonical ensemble) or grand potential (grand canonical ensemble) w.r.t. $\gamma$ under the ensemble $N$-representability constraints $\{0 \leq n_p \leq 1\}$.

If we describe our system in the canonical ensemble, with fixed particle number, we have $\mathcal{T} = \mathcal{H}_N$, where $\mathcal{H}_N$ is the $N$-particle Hilbert space. In the grand canonical ensemble, we have instead $\mathcal{T} = \mathcal{F}$, where $\mathcal{F} = \bigoplus_{N=0}^\infty \mathcal{H}_N$ is the Fock space. Note that in both cases the domain of the functional is different, in the canonical ensemble the functional is only defined for fixed $\mathrm{Tr}(\gamma) = N$, where $N$ is an integer, while for the grand canonical ensemble there is no such constraint. Since $\mathcal{H}_N \subset \mathcal{F}$ we obtain directly from the variational principle $\FE^\beta_{\mathcal{H}_N}[\gamma] \geq \FE^\beta_{\mathcal{F}}[\gamma]$.

The introduction of a non-interacting reference system with the same 1-RDM as the interacting system is also highly desirable as evidenced from the success of the introduction of a single determinant reference in Kohn-Sham Density Functional Theory. The 1-RDM of the interacting system is not idempotent and therefore, unlike the electron density, cannot be obtained from a single determinant reference and one must resort to one of many non-interacting ensembles that yield the correct 1-RDM. At finite-temperature this indeterminacy can be resolved by approximating $\FE^\beta_\mathrm{int}[\gamma]$ with the following non-interacting free energy functional $\FE_0^\beta[\gamma]$,
\begin{align}
\FE_0^\beta[\gamma] &:= \min_{\hat{\Gamma}_0 \rightarrow \gamma} \left(- \frac{1}{\beta} S[\Gamma_0]\right) =  \min_{\hat{\Gamma}_0 \rightarrow \gamma} \left(\frac{1}{\beta} \Tr(\hat{\Gamma}_0 \log(\hat{\Gamma}_0)\right) \nonumber\\
&= - \frac{1}{\beta} \max_{\hat{\Gamma}_0 \rightarrow \gamma} \left(-\mathrm{Tr}(\hat{\Gamma}_0 \log(\hat{\Gamma}_0)\right) =: - \frac{1}{\beta} S_0[\gamma], \label{eq:nonintfreeenergy}
\end{align}
where $\beta = \frac{1}{k_B T}$ and $S_0[\gamma]$ is the non-interacting entropy. Note that the non-interacting entropy is independent of temperature. It is easy to show as well that $S_0[\gamma]$ only depends on the eigenvalues, the Natural Orbital Occupation Numbers (NOONs) $\{n_p\}$, of the 1-RDM. Therefore we can write $S_0[\{n_p\}]$ and $\FE_0^\beta[\{n_p\}]$ instead. A derivation of $S_0[\{n_p\}]$ by a Lagrangian approach is given in appendix \ref{app:deriventropy}.

The minimizing $\hat{\Gamma}_0$ is an ensemble that has as eigenstates the Slater Permanents/Determinants constructed from the eigenvectors of the 1-RDM, the Natural Orbitals (NOs). It is one particular realization of the non-interacting ensemble yielding the correct 1-RDM used in the original proof of the ensemble $N$-representability of the 1-RDM by Coleman\cite{Col-RMP-63}, which shows that if the NOONS $\{n_p\}$ all satisfy $0 \leq n_p (\leq 1)$ we have bosonic (fermionic) ensemble $N$-representability. At finite temperature occupation numbers instead satisfy $0 < n_p (<1)$.\cite{GieRug-PR-19} Dual to the occupation numbers of the 1-RDM are the effective orbital energies $\{ \epsilon_p^\beta \}$, which combine in the non-interacting Hamiltonian $\hat{H}_0^\beta = \sum_p \epsilon_p^\beta a^\dagger_p a_p$, such that in the canonical ensemble,
\begin{equation}
\hat{\Gamma}_0 = \frac{e^{- \beta \hat{H}_0^\beta}}{\mathrm{Tr}(e^{- \beta \hat{H}_0^\beta)})},
\end{equation}
while in the grand canonical ensemble, the effective orbital energies $\{ \epsilon_p^{\beta, \mu} \}$ combine in the non-interacting Hamiltonian $\hat{H}_0^{\beta, \mu}$,
\begin{equation}
\hat{\Gamma}_0 = \frac{e^{- \beta (\hat{H}_0^{\beta, \mu}-\mu \hat{N})}}{\mathrm{Tr}(e^{- \beta {\hat{H}_0^{\beta, \mu}- \mu \hat{N})}})}.
\end{equation}
Evaluating the interaction energy on $\hat{\Gamma}_0$ leads us to a zeroth-order approximation to the interaction energy, $W_0[\gamma] = \mathrm{Tr}(\hat{\Gamma}_0 \hat{W})$. What then remains to be approximated is the correlation free energy $\FE_c^\beta[\gamma]$,
\begin{equation}
\FE_c^\beta[\gamma] = \FE^\beta_\mathrm{int}[\gamma] - W_0[\gamma] + \frac{1}{\beta} S_0[\{n_p\}],
\end{equation}
which can be decomposed into $W_c^\beta[\gamma] = W^\beta[\gamma] - W_0[\gamma]$ and $S_c^\beta[\gamma] = S^\beta[\gamma] - S_0[\{n_p\}]$, such that $\FE_c^\beta[\gamma] = W_c^\beta[\gamma] - \frac{1}{\beta} S_c^\beta[\gamma]$. $\FE_c^\beta[\gamma]$ then needs to be approximated, but this is beyond the scope of this work.

As of yet, the temperature-dependent 1-RDMFT formalism has always been introduced within the grand-canonical ensemble.\cite{BalCanGro-PRA-15, GieRug-PR-19} This follows the general trend that the physics of non-interacting, or mean-field approximations to interacting, bosonic and fermionic quantum systems at non-zero temperature are typically studied within the grand canonical ensemble, even if there is little to no particle exchange at the corresponding temperature and therefore the canonical ensemble is sufficient for an accurate description of the system. 

The reasons for this are also of a practical nature: whereas closed-form expressions exist for expectation values of grand canonical ensembles of non-interacting bosons and fermions there are no such simple expressions for the canonical ensemble. However, in the treatment of interacting systems and their approximations by a mean-field, the canonical ensemble may carry an advantage, if only because the $N$-particle Hilbert space $\mathcal{H}_N$ is in general much smaller than the Fock space $\mathcal{F}$. 

For this reason it is interesting to study the canonical ensemble within temperature-dependent 1-RDMFT and develop numerically stable algorithms to be able to perform calculations on sizeable systems. The basic ingredient of these algorithms is the computation of non-interacting bosonic and fermionic partition functions to obtain expectation values given a set of orbital energies $\{ \epsilon_p \}$ at a certain temperature $\beta$. A recursive method to calculate the canonical partition function for non-interacting systems was described by Borrmann and Franke\cite{BorFra-JCP-93}, and has been applied to compute expectation values for bosonic and simple fermionic systems.\cite{Sch-PRA-17, BarYuMae-PRR-20}

Of particular relevance for 1-RDMFT are computing the non-interacting entropy $S_0[\{n_p\}]$, and the one-body $\langle \hat{n}_p \rangle$ and two-body $\langle \hat{n}_p \hat{n}_q \rangle$ expectation values of the corresponding non-interacting ensemble. In the case of the grand canonical ensemble these expressions are analytical, given by the Bose-Einstein and Fermi-Dirac distribution (upper sign for bosons, lower sign for fermions),
\begin{align}
S_{0,\mathcal{F}}[\{n_p\}] &= - \sum_p n_p \log(n_p) - \sum_p (1\pm n_p) \log(1\pm n_p) \label{eq:SGC}\\
n_p &:= \langle \hat{n}_p \rangle = \frac{1}{e^{\beta (\epsilon_p - \mu)} \pm  1}, \label{eq:occGC}\\
\langle \hat{n}_p \hat{n}_q \rangle &= \langle \hat{n}_p \rangle \langle \hat{n}_q \rangle = n_p n_q,\\
W_{0,\mathcal{F}}[\gamma] &:= \frac{1}{2} \sum_{pq} (n_p n_q- \delta_{pq} n_p) \langle pq || pq \rangle_{\pm},
\end{align}
where $\{ \langle pq|| pq \rangle_{\pm} \}$ are the (anti-)symmetrized integrals of the two-particle interaction in the NO basis. Note that $\lim_{n_p \downarrow 0}\frac{\partial S_0[\{n_p\}]}{\partial n_p}$ diverges, and for fermions also $\lim_{n_p \uparrow 1}\frac{\partial S_0[\{n_p\}]}{\partial n_p}$ diverges. This greatly simplifies the optimization over the 1-RDM, because it removes the necessity of enforcing the Karush-Kuhn-Tucker (KKT) conditions, and provides an important motivation to utilize a non-interacting reference system, even at zero temperature.

The development of new 1-RDM functionals is greatly enhanced by a straightforward and efficient method to invert the relation between the natural orbital occupation numbers $\{ n_p \}$ and the effective orbital energies $\{ \epsilon_p \}$. This is especially true for the development of functionals that heavily utilize data, e.g. functionals based on machine learning. This is also true for Kohn-Sham Density Functional Theory (KS-DFT), where attempts at machine-learning the exchange-correlation functional have used the Kohn-Sham orbitals of approximate functionals instead of the exact Kohn-Sham orbitals.\cite{KirMcMTur-Science-21} Inversion in KS-DFT to obtain the exact Kohn-Sham orbitals and Kohn-Sham potential has therefore seen a great deal of interest, because of its ability to provide insight in the behaviour of the exact functional and elucidate failures of approximate functionals. However, an important complication in KS-DFT does not occur in 1-RDMFT: if one performs a full Configuration Interaction (CI) calculation with a particular finite one-particle basis, one can in general not obtain the same electron density with Kohn-Sham orbitals expressed in the same one-particle basis without introducing fractional occupations.\cite{GieRug-PR-19,OspSta-JCTC-18} 

Returning to 1-RDMFT in the grand canonical ensemble, equation \ref{eq:occGC} is easily inverted to give,
\begin{equation}
\epsilon_p = - \frac{1}{\beta} \log(\frac{n_p}{1 \pm n_p}) + \mu, \label{eq:GCinversion}
\end{equation}
but no such direct inversion is possible in the canonical ensemble. In this work we represent a simple, efficient and robust procedure to obtain the non-interacting potential via a bosonic/fermionic Sinkhorn algorithm, which differs from the usual Sinkhorn\cite{Sin-CJM-66,Cut-ANIPS-13} algorithm in that the bosonic/fermionic symmetry is taken into account explicitly.

As a final note regarding the theoretical aspects, there is the issue of the so-called non-interacting $v$-representability (in this work $v= h$) of the 1-RDM $\gamma$ with occupation numbers $0 < n_p < 1 \, \forall p$. That is, whether or not there exists a set of orbital energies $\{ \epsilon_r\}$ such that
\begin{equation}
\frac{\mathrm{Tr}(a^\dagger_{p} a_q e^{- \beta \sum_r \epsilon_r a^\dagger_r a_r}}{\mathrm{Tr}(e^{- \beta \sum_r \epsilon_r a^\dagger_r a_r})} = \delta_{pq} n_p.
\end{equation} 
Recent work within the grand canonical ensemble has shown that at any interaction strength such an 1-RDM is $v$-representable, and furthermore the corresponding potential/effective orbital energies are unique.\cite{GieRug-PR-19} Similar results have been obtained within the canonical ensemble, but have not yet been published.\cite{SutGie-unpublished-22} In the following we will assume canonical non-interacting $v$-representability, and find indeed for several examples that it holds at least to numerical accuracy.

This work is organized in the following way, first we introduce the Bosonic and Fermionic Sinkhorn algorithms in section \ref{sec:bfsinkhorn}, then we discuss various numerical considerations that come in to play in section \ref{sec:numerical}. Section \ref{sec:implementation} discusses the implementation of the algorithms in the \verb|bfsinkhorn| package. The algorithms are tested in section \ref{sec:results} and a conclusion is given in section \ref{sec:conclusions}

\section{Bosonic and Fermionic Sinkhorn algorithms}
\label{sec:bfsinkhorn}
Computing the canonical partition function and the occupation numbers given a set of orbital energies is more complicated than in the grand canonical ensemble. We  will need to compute so-called auxiliary partition functions, which are introduced below, to compute occupation numbers. See the work of Barghati et al.\cite{BarYuMae-PRR-20} for the introduction of the concept of auxiliary partition functions and a detailed discussion. We will now separately treat first fermions and then bosons. We start with the following expression for the occupation numbers for fermions,
\begin{equation}
n_p = \frac{e^{-\beta \epsilon_p} Z^{\setminus p}_{-, N-1}}{Z_{-,N}}, \label{eq:expnF}
\end{equation}
where $Z_{-,N}$ is the partition function of the $N$-fermion system and with $Z^{\setminus p}_{N-1}$ we denote the auxiliary partition function, which corresponds to the $N-1$ particle system in which the orbital $p$ has been removed. From the definition of the partition function for fermions $Z_{-, N}$ we find that,
\begin{equation}
Z_{-,N} = Z^{\setminus p}_{-,N} + e^{-\beta \epsilon_p} Z^{\setminus p}_{-, N-1}.
\end{equation}
We rewrite the expression for $n_p$ and isolate $e^{-\beta \epsilon_p}$ on the left,
\begin{equation}
e^{- \beta \epsilon_p } = \frac{n_p}{1-n_p} \frac{Z^{\setminus p}_{-, N}}{Z^{\setminus p}_{-, N-1}},
\end{equation}
where now the right hand side is independent of $\epsilon_p$. Taking the logarithm, we obtain the following expression for $\epsilon_p$,
\begin{equation}
\epsilon_p = - \frac{1}{\beta} \log(\frac{n_p}{1-n_p}) - \frac{1}{\beta} \log(Z^{\setminus p}_{-, N}) + \frac{1}{\beta} \log(Z^{\setminus p}_{-, N-1}). \label{eq:update1fermion}
\end{equation}
In the bosonic case the occupation numbers are given instead by,
\begin{equation}
n_p = \frac{e^{-\beta \epsilon_p} Z^{\cup p}_{N-1}}{Z_N}, \label{eq:expnB}
\end{equation}
here $Z^{\cup p}_{+,N-1}$ denotes the auxiliary partition function corresponding to a system of $N-1$ particles with an \textit{extra} energy level with energy equal to $\epsilon_p$ added. For the bosonic partition function $Z_{+, N}$ it holds that,
\begin{equation}
Z_{+,N} = Z^{\cup p}_{+,N} - e^{- \beta \epsilon_p} Z^{\cup p}_{+,N-1}.
\end{equation}
From these relations the following expression for $\epsilon_p$ can be derived,
\begin{equation}
\epsilon_p = - \frac{1}{\beta} \log(\frac{n_p}{1+n_p}) - \frac{1}{\beta} \log(Z^{\cup p}_{+,N}) + \frac{1}{\beta} \log(Z^{\cup p}_{+,N-1}) . \label{eq:update1boson}
\end{equation}
The first terms of equations \ref{eq:update1fermion} and \ref{eq:update1boson} are identical to the corresponding fermionic/bosonic expression for the potential in the grand canonical ensemble (equation \ref{eq:GCinversion}). The last two terms can be seen as a ``correction" for the canonical ensemble.

In practice we will often not work directly with the (auxiliary) partition functions for numerical stability reasons. Instead, we utilize the corresponding (auxiliary) free energies $\FE = - \frac{1}{\beta}\log(Z)$. The advantage from using the free energy arises from the fact that the partition function scales roughly as $N!$, while the free energy via Stirling's approximation scales roughly as $N \log(N)$. Note that in terms of free energies $n_p = e^{-\beta \epsilon_p - \beta(\FE_{\pm, N-1}^{\cup p/\setminus p}-\FE_{\pm, N})}$. We propose to then iterate equation \ref{eq:update1boson} or \ref{eq:update1fermion} to convergence, with the r.h.s. computed from the current $\epsilon_p^{(i)}$ and the l.h.s. the updated $\epsilon_p^{(i+1)}$. In this process we repeatedly need to compute for a given set $\{\epsilon_p \}$ the (auxiliary) partition functions free energies for the system with particle numbers $M$ ranging from $0$ to $N$. The free energies are obtained from the expression of Borrmann and Franke,\cite{BorFra-JCP-93}
\begin{align}
C_k &= \sum_p e^{- \beta \epsilon_p k},\\
Z_{\pm, M} &= \frac{1}{M}\sum_{k=1}^M (\pm)^{k-1} C_k Z_{\pm, M-k} \label{eq:BorrmannFranke}.
\end{align}
In both cases we instead compute the corresponding free energies,
\begin{align}
D_k &= - \frac{1}{\beta}\log(C_k),\\
\FE_{\pm, M} &= -\frac{1}{\beta} \log\left(\sum_{k=1}(\pm)^{k-1} e^{- \beta D_k - \beta \FE_{\pm, M-k}}\right) + \frac{1}{\beta} \log(M). \label{eq:freeenergy}
\end{align}
The auxiliary partition functions/free energies for particle number $M$ can be obtained from the partition functions/free energies for particle numbers ranging from $0$ to $M$,\cite{BarYuMae-PRR-20}
\begin{align}
Z_{\pm,M}^{\cup /\setminus p} &= \sum_{k=0}^M (\pm)^{k} e^{- \beta \epsilon_p k} Z_{\pm, M-k},\\
\FE_{\pm, M}^{\cup /\setminus p} &= -\frac{1}{\beta} \log \left(\sum_{k=0}^{M} (\pm)^k e^{- \beta \epsilon_p k - \beta \FE_{\pm, N-k}^p} \right). \label{eq:auxillary}
\end{align}

\begin{figure*}
    \includegraphics{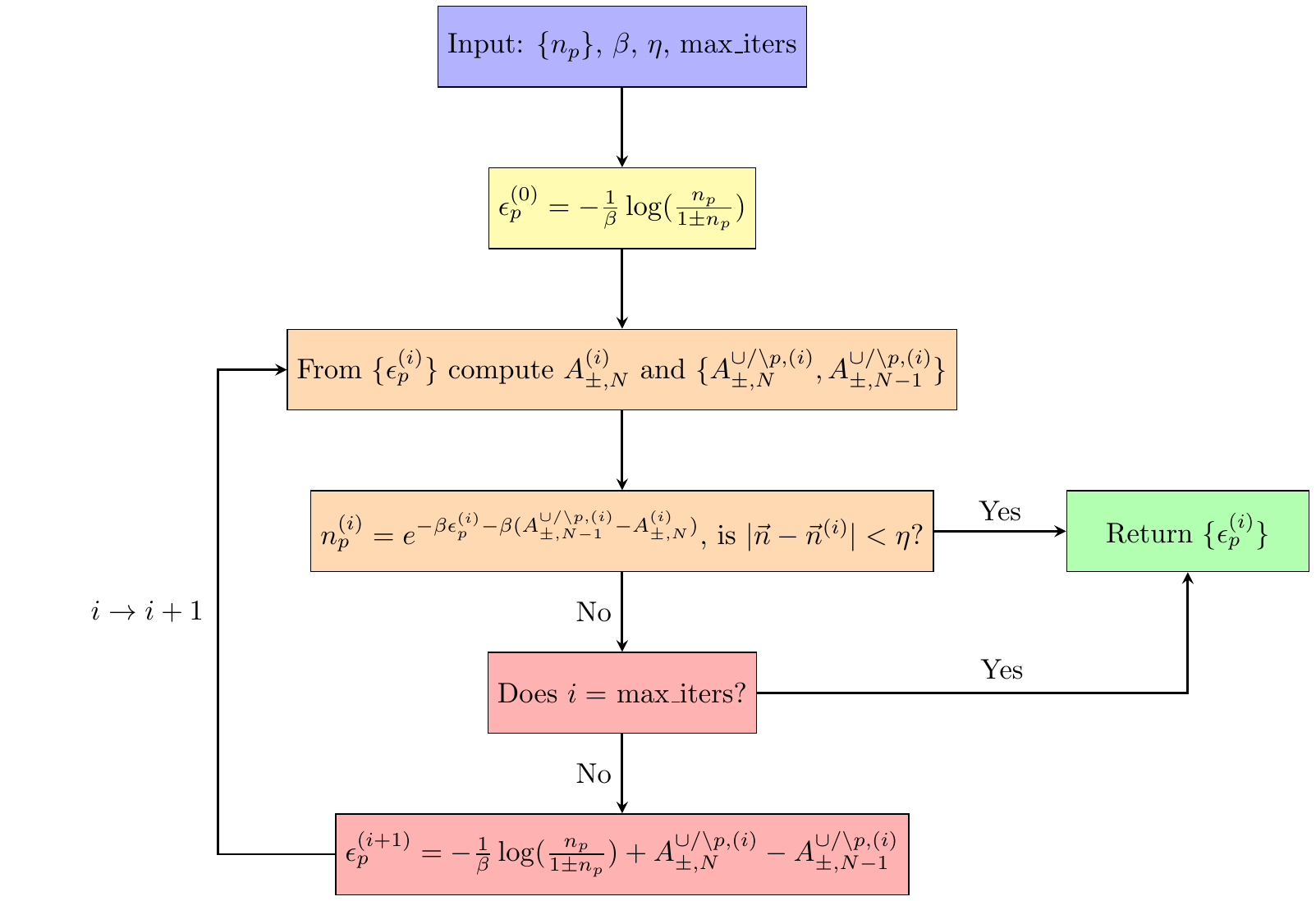}
    \caption{Block scheme of the Bosonic and Fermionic Sinkhorn algorithms. $\{n_p\}$ are the NOONs. $\beta$ is the inverse temperature. $\eta$ is the tolerance of the errors in the NOONs. max\_iters is the maximum number of iterations. $\{ \epsilon_p^{(i)} \}$ are the NO energies at iteration $i$. $\FE^{(i)}_{\pm,N}$ is the bosonic/fermionic non-interacting free energy at iteration $i$. $\FE^{\cup/\setminus p, (i)}_{\pm,N}$ and $\FE^{\cup/\setminus p, (i)}_{\pm,N-1}$ are the bosonic/fermionic non-interacting auxiliary free energies for the system with, respectively, $N$ and $N-1$ particles at iteration $i$. }
    \label{fig:algorithm}
\end{figure*}

We start our calculation from given input NOONs $\{n_p \}$, inverse temperature $\beta$, a tolerance $\eta$ and a maximum number of iterations. NOONs that are numerically close to zero (or one for fermions) are removed from the calculation for numerical stability reasons. This does not affect the final entropy, because states containing NOs with NOONs close to zero have near zero weight, while in the fermionic case all states with non-zero weight will contain the NO with NOON close to 1. We use as a cut-off for NOONS a lower limit of $10^{-12}$ and an upper limit of $1-10^{-12}$ for fermions. We label our quantities with iteration number $i$, where $i=0$ corresponds to the starting guess. The steps of the full algorithm are shown in figure \ref{fig:algorithm}. Note that we obtain a starting guess from the corresponding expression in the grand canonical ensemble (equation \ref{eq:GCinversion}).

The computational cost of $\FE_{\pm, N}$ scales quadratically for large $N$ and needs to be performed $N_\mathrm{iter}$ times for an overall scaling of $N^2 N_\mathrm{iter}$. The computation of $\FE_{\pm, M}^{\cup/\setminus p}$ scales linearly in $N$, but needs to be performed for $N_\mathrm{orb}$ NOs and also needs to be performed $N_\mathrm{iter}$ times for an overall scaling of $N N_\mathrm{orb} N_\mathrm{iter}$. The calculation of $\FE_{\pm, M}^{\cup/\setminus p}$ is done in a way that is embarrassingly parallel in the orbitals. If we then take $N_\mathrm{orb} \propto N$ (as is usually the case) and assume that the $N_\mathrm{iter}$ is independent of $N$ then we obtain an overall computational cost scaling as $N^2$. Since the algorithm only involves rather elementary steps, in particular computing exponentials and logarithms, the prefactor is also very small.

We note that this algorithm (we will refer to it as ``Bosonic/Fermionic Sinkhorn") is closely related to the Sinkhorn algorithm which results from entropically regularized Multi Marginal Optimal Transport (MMOT) \cite{BenCarCut-SJSC-15,BenCarNen-SMCISCE-16}, which has been applied to the strongly-interacting Limit of DFT within the Strictly Correlated Electrons (SCE) formalism.\cite{ButPasGor-PRA-12,MarGerNen-TOOT-17,GerGroGor-JCTC-19}

The ``particle statistics naive" Sinkhorn algorithm (which we will simply refer to as Sinkhorn) in this case corresponds to the direct iteration of equation \ref{eq:expnB} and \ref{eq:expnF}, which yields,
\begin{equation}
\epsilon_p^{(i+1)} = - \frac{1}{\beta} \log(n_p) + \FE_{N}^{(i)} - \FE^{ p (i)}_{N-1}, \label{eq:Sinkhorn}
\end{equation}
which is for fermions identical to the application of the usual Sinkhorn algorithm with a pairwise symmetric cost given by,
\begin{equation}
c_{pq} = \begin{cases} \infty & p = q\\
0 & p \neq q.
\end{cases}
\end{equation}

For bosons, there is no such easy mapping, due to the fact that in MMOT the particles are assumed to be distinguishable and therefore an overcounting of the number of states arises. An in-depth discussion of the connection to MMOT and how particle statistics can be recovered via an effective cost is the subject of appendix B. 

The non-interacting entropy and free energy can be obtained from the converged $\{ \epsilon_p^\beta \}$ and the corresponding corresponding free energy $\FE_N^\beta$ via,
\begin{align}
S_0[\{n_p\}] &= \beta \left( \sum_p n_p \epsilon_p^\beta - \FE_N^\beta \right),\\
\FE_0^\beta[\{n_p\}] &= \FE_N^\beta - \sum_p n_p \epsilon_p^\beta.
\end{align}
Since the occupation numbers and indeed all expectation values are invariant under a constant shift of all the orbital energies ($\epsilon_p \rightarrow \epsilon_p + C$) we are free to make a suitable choice. It is physically elegant to fix the arbitrary constant in our potential such that,
\begin{equation}
\FE_0^\beta[\{n_p\}] = \FE_N^\beta,
\end{equation}
and so we can directly identify the non-interacting free energy with the free energy of the non-interacting ensemble. This is achieved by shifting the orbital energies by a constant such that $\sum_p n_p \epsilon_p^\beta = 0$. In the following we will always enforce this condition at every iteration, but in practical applications other choices may be possible. In KS-DFT one often desires that the Kohn-Sham potential for a finite system in the spatial representation vanishes infinitely far from the system, fixing the gauge of the potential, but in this case such a criterion is not immediately accessible. Also in KS-DFT other choices of the gauge are available, e.g. such that the energy of the Kohn-Sham system is identical to that of the interacting system.\cite{VucLevGor-JCP-17} 

Given the converged $\{\epsilon_p \}$ we can also obtain the approximation $W_0[\gamma]$ to the interaction energy via the two-particle correlations,\cite{BarYuMae-PRR-20}
\begin{align}
\langle \hat{n}_p \hat{n}_q \rangle &= \mp \frac{e^{\beta \epsilon_p^\beta[\{n_p\}]}n_p - e^{\beta \epsilon_q^\beta[\{n_p\}]}n_q}{e^{\beta \epsilon_p^\beta[\{n_p\}]}- e^{\beta \epsilon_p^\beta[\{n_p\}]}} \quad \forall p \neq q, \label{eq:2correl}\\
\langle \hat{n}_p^2 \rangle_+ &= \frac{1}{Z_{+,N}}\sum_{k=1}^N (2 k - 1) e^{- \beta \epsilon_p k} Z_{+,N-k} \\
\langle \hat{n}_p^2 \rangle_- &= n_p \\
W_0[\gamma] &= \frac{1}{2 }\sum_{pq} (\langle \hat{n}_p \hat{n}_q \rangle - \delta_{pq} n_p)  \langle pq || pq \rangle_{\pm}, \label{eq:interactionenergy}
\end{align}
where the expression for $W_0[\gamma]$ follows from $\langle \hat{n}_p \hat{n}_q \rangle = \langle \hat{a}_p^\dagger \hat{a}_p \hat{a}^\dagger_q \hat{a}_q \rangle = \pm \langle \hat{a}_p^\dagger \hat{a}^\dagger_q  \hat{a}_p \hat{a}_q \rangle + \delta_{pq} \langle \hat{a}_p^\dagger \hat{a}_p \rangle = \langle \hat{a}_p^\dagger \hat{a}^\dagger_q  \hat{a}_q \hat{a}_p \rangle + \delta_{pq} n_p$. Because $W_0[\gamma]$ depends implicitly on $\{n_p\}$ via $\{\epsilon_p \}$ self-consistent optimization requires computing the derivatives $\{ \frac{\partial \epsilon_p}{\partial n_q}\}$.

For degenerate orbitals equation \ref{eq:2correl} is ill-defined as both the numerator and denominator become zero. In this case we use equation 48 Barghati et al.\cite{BarYuMae-PRR-20} for the case of degeneracy,
\begin{equation}
\langle \hat{n}_p \hat{n}_q \rangle = \frac{1}{Z_{\pm, N}}\sum_{k=2}^N (\pm)^{k} (k-1) e^{- \beta \epsilon_p k} Z_{\pm, N-k} \quad \mathrm{if}\quad \epsilon_p = \epsilon_q,
\end{equation}
which must be implemented with an appropriate method of determining degeneracy. In this work we use the criterion $|e^{\beta \epsilon_p} - e^{\beta \epsilon_q}| < 10^{-8}$, which corresponds to the denominator of equation \ref{eq:2correl}. A more general implementation in which one does not distinguish degenerate and non-degenerate can be obtained from the more general expression,
\begin{equation}
\langle \hat{n}_p \hat{n}_q \rangle = \frac{e^{- \beta (\epsilon_p+\epsilon_q)} Z_{\pm, N-2}^{\cup/\setminus pq}}{Z_{\pm, N}}.
\end{equation}

\section{Numerical considerations}
\label{sec:numerical}
Note that our inversion algorithm depends on the choice of $\beta$, but for every $\beta$ the same ensemble $\hat{\Gamma}$ is found and therefore also the same value for $S_0[\{n_p\}]$ and $W_0[\gamma]$. The converged $\{\epsilon_p[\{n_q\}]\}$ satisfy a simple scaling relation in $\beta$,
\begin{equation}
\epsilon^\beta_p[\{n_q\}] = \frac{\beta_\mathrm{inv}\epsilon^{\beta_\mathrm{inv}}_p[\{n_q\}]}{\beta},
\end{equation}
where $\beta_\mathrm{inv}$ can be chosen for numerical stability. We find however that the algorithm is mostly insensitive to this choice and therefore we set $\beta_\mathrm{inv} = 1$ in the following. In the bosonic case, we can compute the required (auxiliary) free energies by stabilizing the logarithm of a sum of exponential terms in the following way (a ``log-sum-exp'' trick),
\begin{equation}
- \frac{1}{\beta} \log(\sum_i e^{- \beta a_i}) = - \frac{1}{\beta} \log\left(\sum_i e^{- \beta (a_i-a_\mathrm{min})}\right) + a_\mathrm{min},
\end{equation}
which avoids large exponents in the log-sum. In the fermionic case the ``log-sum-exp'' trick is less effective, because of the signs present in the summation of equations \ref{eq:freeenergy} and \ref{eq:auxillary}. A simple way of avoiding this issue is by instead working with the quotient of the partition functions with particle numbers differing by one,
\begin{equation}
Q_M = \frac{Z_M}{Z_{M-1}},
\end{equation}
and the corresponding free energy differences,
\begin{equation}
\Delta \FE_M = - \frac{1}{\beta} \log(Q_M) = \FE_M-\FE_{M+1}.
\end{equation}
Similarly for the quantities $C_M$ and $D_M$ we adopt the notation $R_M = \frac{C_{M}}{C_{M-1}}$ and $\Delta D_M = D_M - D_{M-1}$. Equation \ref{eq:BorrmannFranke} can be then written as a recursion relation in terms of the quotients,
\begin{equation}
Q_M = \frac{1}{M} C_1(1 \pm \frac{R_2}{Q_{M-1}} (1 \pm \frac{R_3}{Q_{M-2}}(1\pm \frac{R_4}{Q_{M-3}}\dots))).
\end{equation}
Equation \ref{eq:auxillary} can be transformed in a similar way.

\section{Implementation}
\label{sec:implementation}
The algorithm was implemented in \verb|python| 3.7.4 with \verb|jax|\cite{jax-google-18} 0.38. The algorithm as well as the code to generate the figures in this work are available as part of the package \verb|bfsinkhorn| here: \href{https://www.github.com/DerkKooi/bfsinkhorn}{https://www.github.com/DerkKooi/bfsinkhorn}. \verb|jax| is used because its ability to perform just-in-time compilation allows for the flexibility of \verb|python|, while retaining good performance. Implementation in \verb|jax| also allows for the use of both forward and reverse automatic differentiation without implementing any additional functions. This allows for the calculation of quantities like $\frac{\partial \epsilon_p}{\partial n_q}$ and arbitrary higher order derivatives. 

\section{Results}
\label{sec:results}
\begin{figure*}
    \centering
    \includegraphics[scale=0.8]{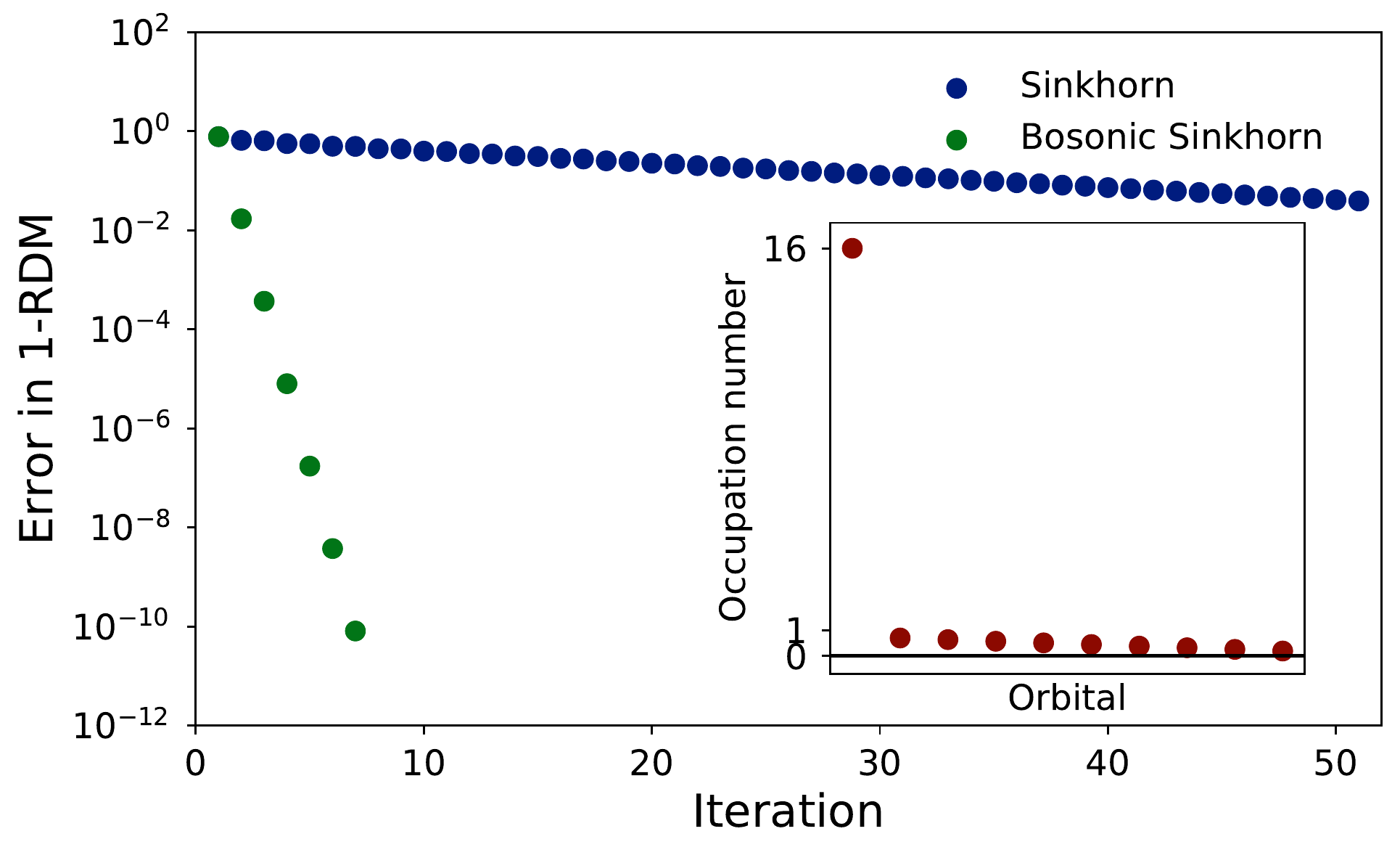}
    \caption{Convergence of the ``naive" Sinkhorn and Bosonic Sinkhorn algorithms to the correct 1-RDM, measured in terms of the 1-norm $|\gamma-\gamma_\mathrm{approx}| = \sum_p |n_p-n_p^\mathrm{approx}|$.
    Inset: the desired (synthetic) distribution of the NOONs. The system has 20 particles in 10 orbitals.}
    \label{fig:boson_convergence}
\end{figure*}

To test the Bosonic Sinkhorn algorithm we have generated synthetic NOON distributions, heavily occupying the first orbital, and then letting the occupation numbers decay. Figure \ref{fig:boson_convergence} shows the convergence of the Sinkhorn and Bosonic Sinkhorn algorithms applied to a particular distribution. In this case, the Bosonic Sinkhorn algorithm converges rapidly, while the Sinkhorn algorithm only converges very slowly. In other cases encountered (see e.g. figure S1 in the supplementary material) the Bosonic Sinkhorn algorithm converges rapidly, while the Sinkhorn algorithm does not converge at all. To demonstrate the scaling of the algorithm, we show convergence for a distribution with $N=1000$ bosons and $N=10000$ orbitals in figure S2 in the supplementary material. These calculations only take seconds on a modern laptop.

\begin{figure*}
    \centering
    \includegraphics[scale=0.8]{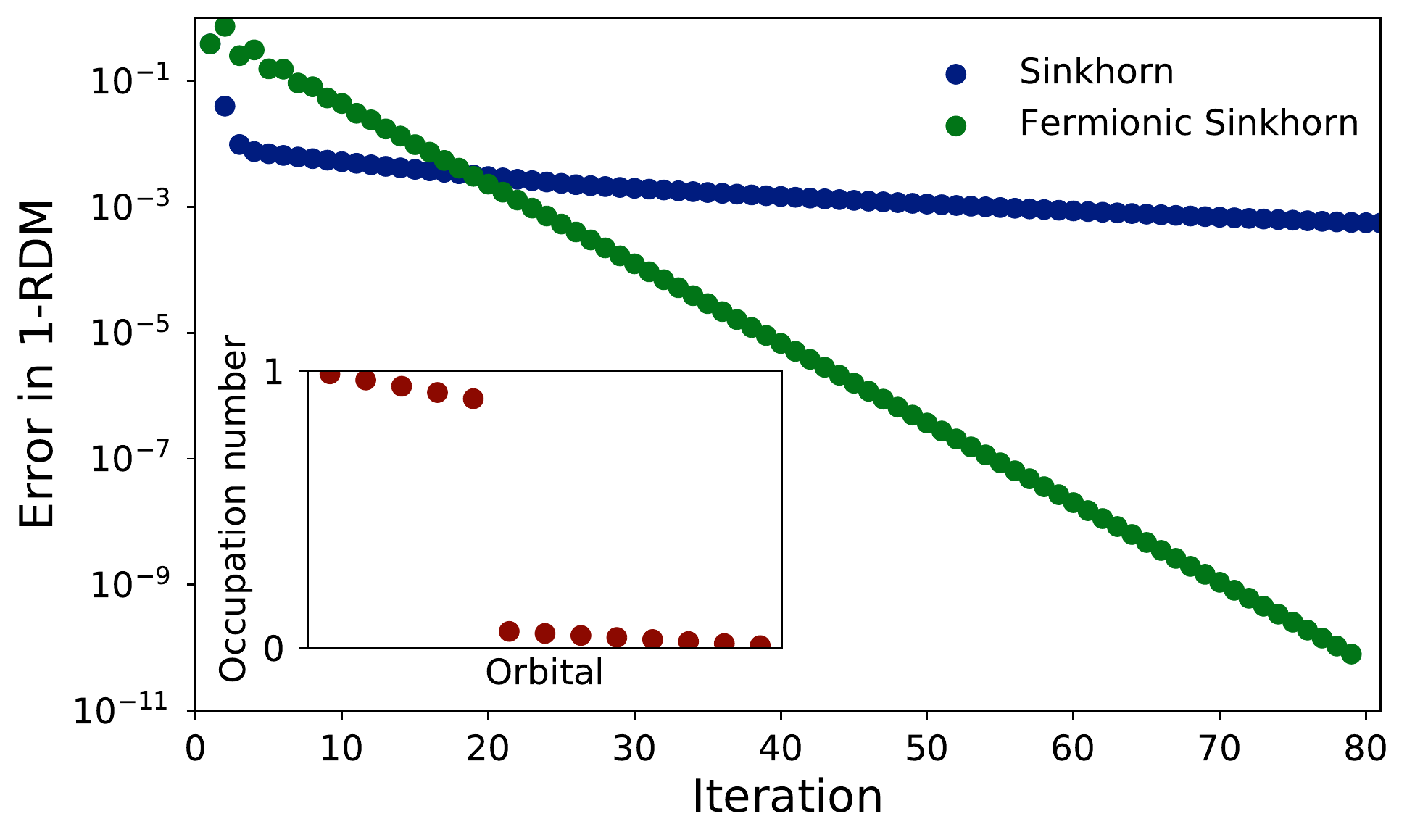}
    \caption{Convergence of the ``naive" Sinkhorn and Fermionic Sinkhorn algorithms to the correct 1-RDM, measured in terms of the 1-norm $|\gamma-\gamma_\mathrm{approx}| = \sum_p |n_p-n_p^\mathrm{approx}|$. Inset: the desired (synthetic) distribution of the NOONs. The system has 5 particles in 13 orbitals. }
    \label{fig:fermion_convergence}
\end{figure*}

For the tests of the Fermionic Sinkhorn algorithm we also generated synthetic NOONs, heavily occupying the first $N$ orbitals in a decaying manner, and then occupying the remaining orbitals again with a decay. Figure \ref{fig:fermion_convergence} shows the convergence of the Sinkhorn and Fermionic Sinkhorn algorithm applied to a particular distribution, where the occupations of the first $N$ orbitals are not close to fully occupied. The Fermionic Sinkhorn algorithm initially converges less rapidly than the Sinkhorn algorithm, but eventually converges very close to the 1-RDM, while the Sinkhorn algorithm does not. In the following, we will therefore always start with 10 iterations of Sinkhorn before starting the Fermionic Sinkhorn algorithm to accelerate convergence. The effect of this is illustrated in the supplementary material figure S3.

We will now proceed to apply the Sinkhorn and Fermionic Sinkhorn algorithm to realistic NOON distributions obtained from ground-state electronic calculations using \verb|pyscf|\cite{SunBerBlu-WCMS-17} 2.0.0. In the absence of magnetic fields and neglecting relativistic effects the Hamiltonian commutes with the spin-operators $\hat{S}_z$ and $\hat{S}^2$ and the 1-RDM in that case is block diagonal in spin-up ($\uparrow$) and spin-down ($\downarrow$) orbitals. We can therefore split our NOs in spin-up and spin-down NOs. Our partition function for $N$ electrons can then be decomposed into separate contributions from the different possible occupations $N_\uparrow$ ($N_\downarrow$) of the spin-up (spin-down) NOs as,
\begin{equation}
Z_{N} = \sum_{N_\uparrow=0}^N Z^\uparrow_{N_\uparrow} Z^\downarrow_{N-N_\uparrow},
\end{equation}
such that $N_\uparrow + N_\downarrow = N$. However, we can also choose to further restrict the definition of our free energy functional (equation \ref{eq:functional}) to only include states with a particular expectation value of $\hat{S}_z$ = $\langle \hat{S}_z \rangle = N_\uparrow - N_\downarrow$, which we will denote by the wavefunction subset $\mathcal{S}_z \subseteq \mathcal{H}_N$. Note that this is only possible if $\sum_p n^\uparrow_p = N_\uparrow$ and $\sum_p n^\downarrow_p = N_\downarrow$, with $N_\uparrow$ and $N_\downarrow$ integers. For a state $\langle \hat{S}^2 \rangle = S(S+1)$ we can always choose to work with integer values of $-S \leq \langle \hat{S}_z \rangle \leq S$. In this case our partition function becomes a product of partition functions for spin-up and spin-down,
\begin{equation}
Z_{N, \mathcal{S}_z} = Z_{N_\uparrow}^\uparrow Z_{N_\downarrow}^\downarrow,
\end{equation}
such that
\begin{equation}
\FE^\beta_{0, \mathcal{S}_z}[\{n_p\}] = \FE^\beta_0[\{n^\uparrow_p\}] + \FE^\beta_0[\{n^\downarrow_p\}],
\end{equation}
and we can perform the inversion separately for the spin-up and spin-down NOs as if the corresponding electrons are independent particles.

\begin{figure*}
    \centering
    \includegraphics[scale=0.8]{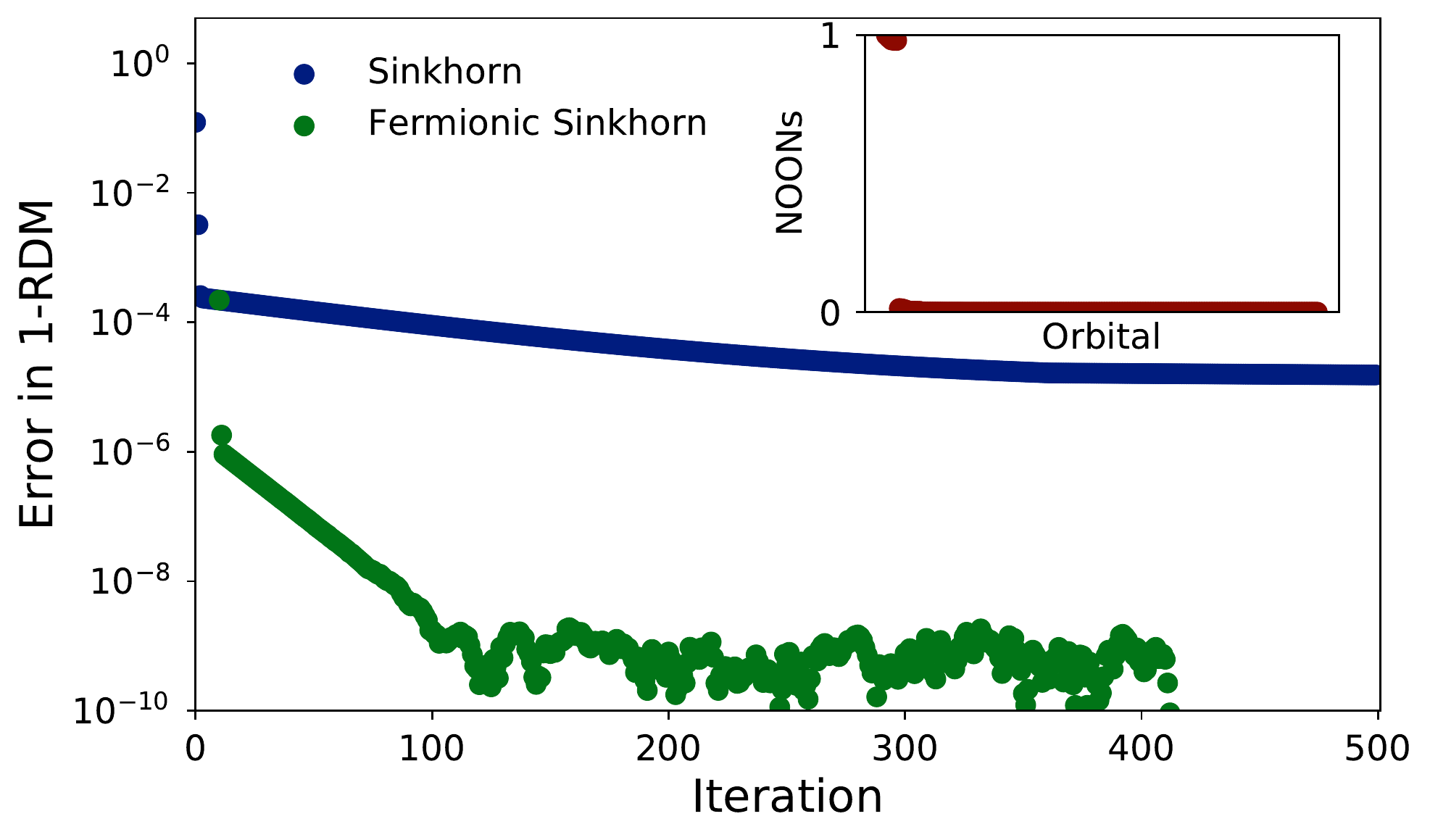}
    \caption{Convergence of the ``naive" Sinkhorn and Fermionic Sinkhorn algorithms of H$_2$O at equilibrium geometry calculated using CCSD in a cc-pVQZ basisset, measured in terms of the 1-norm $|\gamma-\gamma_\mathrm{approx}| = \sum_p |n_p-n_p^\mathrm{approx}|$. The inset shows the distribution of the NOONs.  }
    \label{fig:H2O_convergence}
\end{figure*}

Figure \ref{fig:H2O_convergence} shows the convergence of the Sinkhorn and Fermionic Sinkhorn algorithm within the wavefunction subspace $\mathcal{S}_z$ for singlet H$_2$O at equilibrium geometry obtained from a CCSD calculation in a cc-pVQZ basisset. Again, the Fermionic Sinkhorn algorithm converges much better than the Sinkhorn algorithm, but numerical complications prevent full convergence. Surprisingly, these numerical complications are \textit{worse} if one uses a smaller basis. Having multiple NOONs that have (nearly) the same value seems to play an important role. In the case of exact degeneracy Sinkhorn may converge, while Fermionic Sinkhorn already runs into problems at the first iteration. These issues seem to arise in the computation of the partition functions/free energies and may be resolved by further improvements in their computation.

An additional reduction of the wavefunction space can be obtained by working with the so-called Configuration State Functions (CSFs), which are simultaneous eigenstates of $\hat{S}_z$, $\hat{S}^2$, and the spatial 1-RDM operator $\hat{\tilde{\gamma}}_{pq} = a^\dagger_{p\uparrow} a_{q\uparrow} + a^\dagger_{p\downarrow} a_{q\downarrow}$. We denote the space of Configuration State Functions by $\mathcal{S}$. The expectation value of $\hat{S}^2$ cannot be determined solely from the 1-RDM, therefore we must specify it beforehand and ensure that the 1-RDM fulfills particular ensemble ``spin representability" constraints. That is, given a spatial 1-RDM $\tilde{\gamma}_{pq}$ we must ensure that there exists an ensemble $\hat{\Gamma} \in \mathcal{S} \otimes \mathcal{S}$ such that $\Tr(\hat{\Gamma}\hat{\tilde{\gamma}}_{pq} ) = \tilde{\gamma}_{pq}$.

For singlet states the ``spin representability" conditions are facile: we need an even number of electrons, given $\{0 < \tilde{n}_p < 2 \}$, the eigenvalues of the spatial 1-RDM $\tilde{\gamma}$, we obtain the spin 1-RDM by putting an identical number of electrons in the spin-up and spin-down NOs with the condition $n_{p\uparrow} = n_{p \downarrow} = \frac{\tilde{n}_p}{2} \quad \forall \quad p$. For the $N=2$ singlet the non-interacting problem then becomes identical to the bosonic problem with $N=2$ and NOONs $\{\tilde{n}_p \}$.

\begin{figure*}
    \centering
    \includegraphics[scale=0.8]{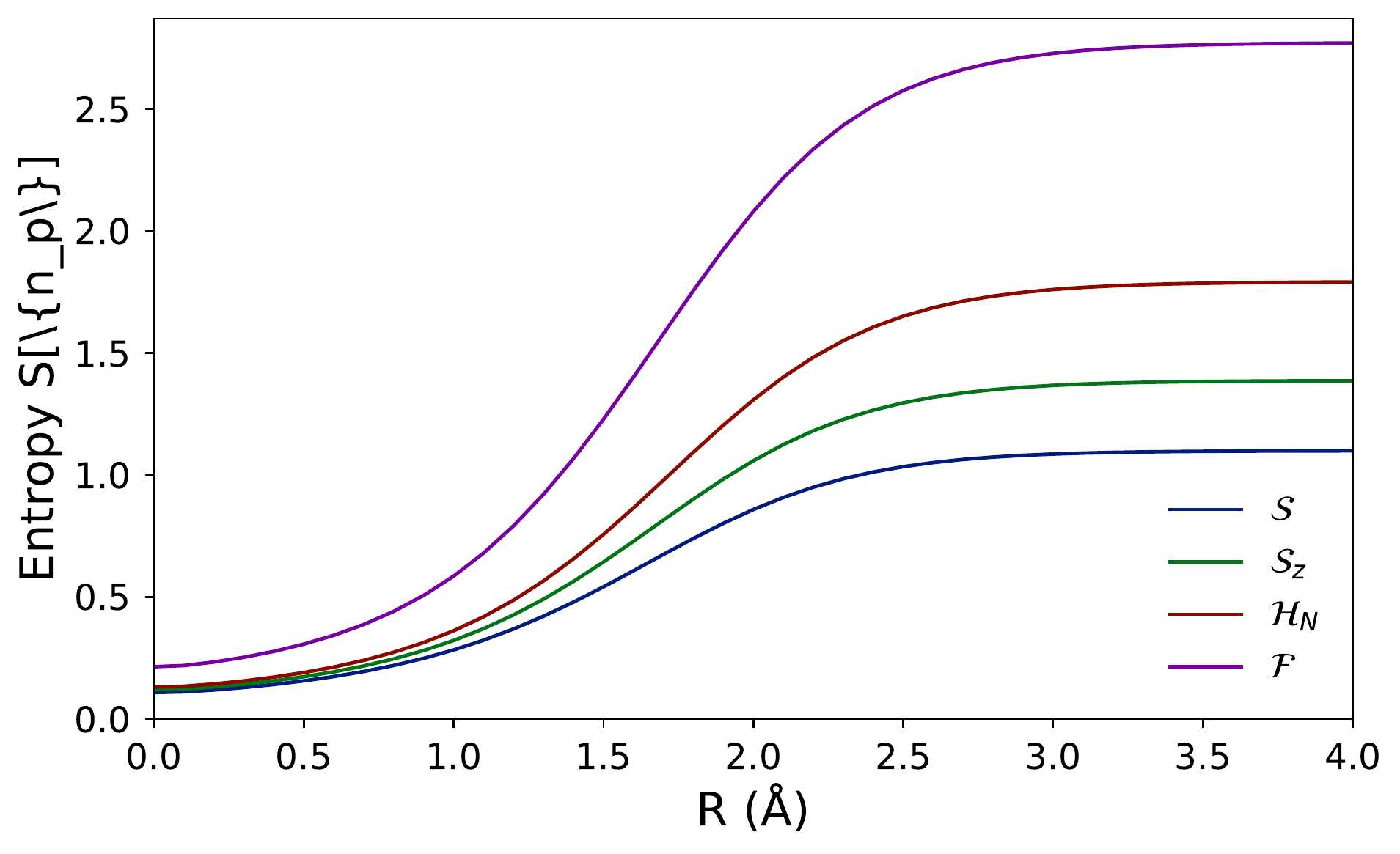}
    \caption{Non-interacting entropy $S_0[\{n_p\}]$ in Hartree atomic units computed for the H$_2$ singlet at different bond lengths $R$ in angstrom in different wavefunction spaces evaluated with the exact (CISD) 1-RDM. }
    \label{fig:H2_singlet_entropy}
\end{figure*}

We illustrate the different choices of wavefunction spaces for the H$_2$ singlet for different bond lengths $R$. The ground state is calculated from CISD in a aug-cc-pVQZ basisset. Figure \ref{fig:H2_singlet_entropy} shows the entropy that is obtained for the different wavefunction spaces. In every wavefunction space the entropy shows similar behaviour: a minimum at $R=0$, then a monotonic increase until saturating for large $R$. The entropies for different wavefunction subspaces show a distinct ordering as is expected since $\mathcal{S} \subseteq \mathcal{S}_z \subset \mathcal{H}_N \subset \mathcal{F}$ and therefore we obtain from the variational principle of the non-interacting free energy $\FE_{0, \mathcal{S}}[\{n_p\}] \geq \FE_{0, \mathcal{S}_z}[\{n_p\}] \geq \FE_{0, \mathcal{S}_z}[\{n_p\}] \geq \FE_{0, \mathcal{H}_N}[\{n_p\}] \geq \FE_{0, \mathcal{F}}[\{n_p\}]$ and so $S_{0, \mathcal{S}}[\{n_p\}] \leq S_{0, \mathcal{S}_z}[\{n_p\}] \leq S_{0, \mathcal{S}_z}[\{n_p\}] \leq S_{0, \mathcal{H}_N}[\{n_p\}] \leq S_{0, \mathcal{F}}[\{n_p\}]$.

\begin{figure*}
    \centering
    \includegraphics[scale=0.8]{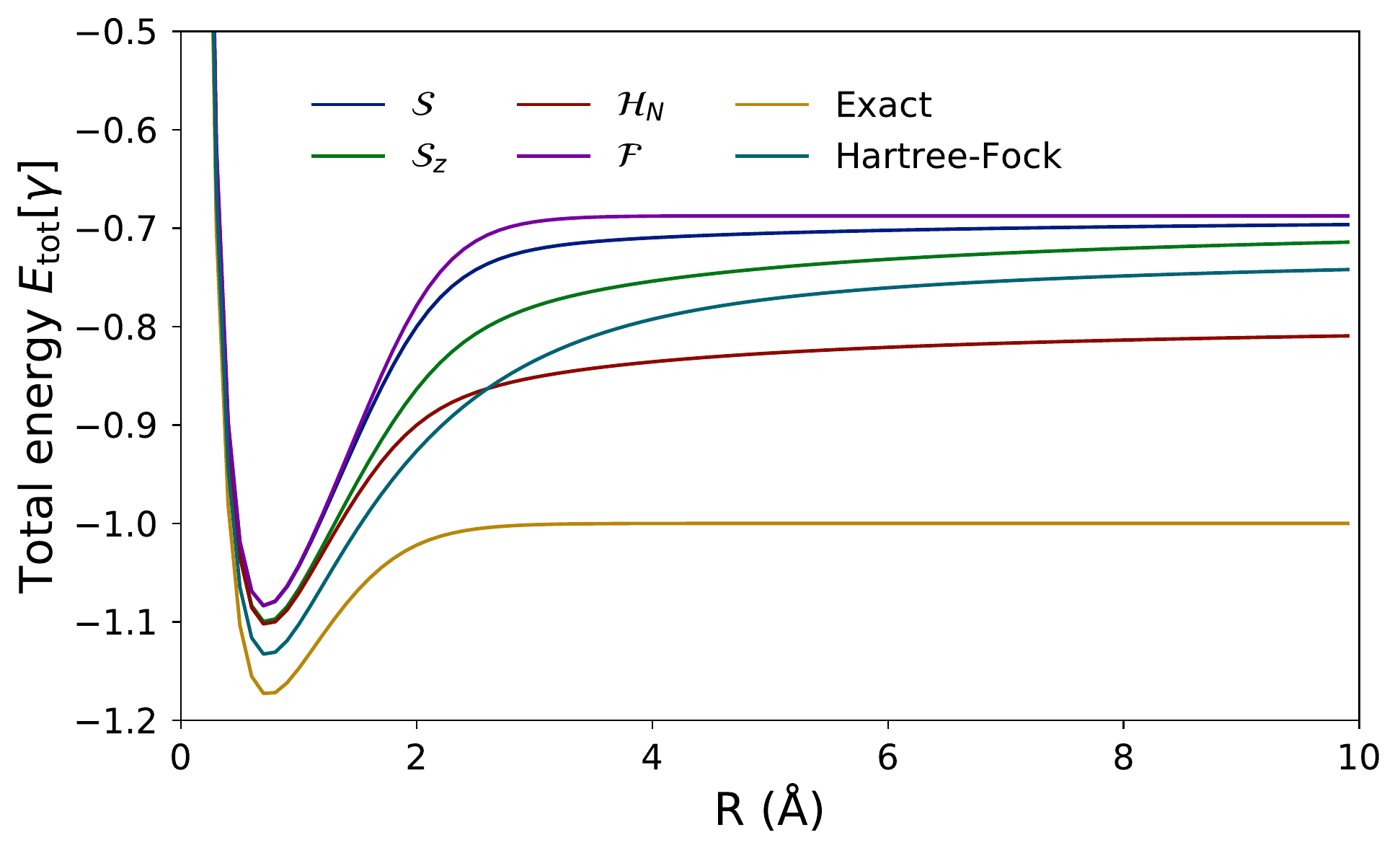}
    \caption{Total energy $E_\mathrm{tot}[\gamma]$ in Hartree computed for the H$_2$ singlet at different bond lengths $R$ in angstrom obtained from CISD (exact), in different wavefunction subspaces evaluated with the exact (CISD) 1-RDM or from Restricted Hartree-Fock. }
    \label{fig:H2_singlet_total_energy}
\end{figure*}

Figure \ref{fig:H2_singlet_total_energy} shows the total energy $E_\mathrm{tot}[\gamma] = h[\gamma] + W[\gamma] = T[\gamma] +V_\mathrm{ext}[\gamma] + W[\gamma]$ obtained from CISD with the exact $W[\gamma]$, and with the approximation $W[\gamma] \approx W_0[\gamma]$ evaluated with the exact CISD 1-RDM for the different choices of spaces. For the explicit expressions of $W_0[\gamma]$ for the different choices of spaces see appendix \ref{app:interactionenergy}. We also compare Restricted Hartree-Fock (RHF), which of course has a different 1-RDM as the exact wavefunction. 

Clearly approximating $W[\gamma]$ with $W_0[\gamma]$ is a crude approximation, since correlation is missing. However, we can still use them as approximations to be improved upon by a correlation functional. $W_{0, \mathcal{F}}[\gamma]$ is conventionally used as the starting point for many zero-temperature 1-RDMFT functionals.\cite{PerGie-DFMES-15} Approximating $W[\gamma] = W_{0, \mathcal{F}}[\gamma]$ in the total energy and then optimizing $\gamma$ has been proven to always yield the HF state.\cite{Lie-PRL-81} Indeed we see that $E_\mathrm{tot}[\gamma]$ obtained with $W_{0, \mathcal{F}}[\gamma]$ is for every $R$ above the RHF state.

We find that using $W_{0, \mathcal{S}}[\gamma]$ is very close to using $W_{0, \mathcal{F}}[\gamma]$, in particular around the equilibrium geometry. Using $W_{0, \mathcal{S}_z}$ and $W_{0, \mathcal{H}_N}$ yield energies slightly below those obtained with $W_{0, \mathcal{S}}[\gamma]$ and $W_{0, \mathcal{F}}[\gamma]$ at equilibrium geometry, but still well above the RHF and exact energies. Dissociation gives a more interesting picture. Using $W_{0, \mathcal{S}_z}[\gamma]$ gives energies well above those obtained from RHF, while using $W_{0, \mathcal{H}_N}[\gamma]$ ends up crossing the RHF energy and giving a significantly smaller error in dissociation. This indicates that in zero-temperature 1-RDMFT $W_{0, \mathcal{H}_N}[\gamma]$ may provide a better starting point for the design of functionals than $W_{0, \mathcal{F}}[\gamma]$. The remainder that still needs to be approximated using a correlation functional $W_c[\gamma]$ is plotted in figure S4 of the supplementary material.

\section{Conclusions}
\label{sec:conclusions}
It is desirable to formulate a finite-temperature 1-body Reduced Density Matrix Functional Theory (1-RDMFT) in the canonical ensemble to describe systems in which particle-number fluctuations are negligible. By introducing the non-interacting (``maximum entropy") approximation and providing an efficient manner of inverting the relationship between occupation numbers and orbital energies we have provided a starting point for this field. The introduction of different ensembles resulting from different choices of wavefunction spaces has also produced several novel non-interacting approximations to the interaction energy $W_0[\gamma]$, which can also be relevant to the development of 1-RDMFT at zero-temperature. At zero temperature the different functionals $S_0[\{n_p\}]$ obtained in the canonical ensemble may also be used in approximations to the correlation energy in a similar fashion as the entropy for the grand canonical ensemble\cite{WanBae-PRL-22}.

The Bosonic and Fermionic Sinkhorn algorithms take into account the particle statistics explicitly and show improved convergence over the ``particle statistics naive" Sinkhorn algorithm. This is especially true for the Bosonic Sinkhorn algorithm, which in all cases studied converges in $\mathcal{O}(10)$ iterations, while Sinkhorn converges slowly or not at all. The Fermionic Sinkhorn algorithm converges more slowly, but still improves over the Sinkhorn algorithm after a certain number of iterations. Starting initially with the Sinkhorn algorithm before switching to the Fermionic Sinkhorn algorithm alleviates the initial bad performance.

The relation between the Bosonic and Fermionic Sinkhorn algorithms and the usual Sinkhorn algorithm used in entropically-regularized Multi-Marginal Optimal Transport may pave the way to finding deeper connections between 1-RDMFT, non-interacting ensembles and MMOT. In particular, the effect of particle number statistics discussed in appendix \ref{app:mmot} may be of relevance to other MMOT problems with identical marginals, while insights from MMOT may provide improved algorithms.

The \verb|bfsinkhorn| package provides the community with a way of rapidly integrating and testing non-interacting functionals based on the canonical ensemble. Many elements of the \verb|bfsinkhorn| package may also be reused for other applications of non-interacting canonical ensembles, even if the Sinkhorn algorithm is not necessary. 

\noindent{\it Acknowledgements --} 
This research was supported by the Netherlands Organisation for Scientific Research (NWO) under Vici grant 724.017.001. The author thanks K.J.H. Giesbertz, P. Gori Giorgi, E.J. Baerends, S.M. Sutter and M. Rodr\'iguez-Mayorga for insightful discussions. The author thanks K.J.H. Giesbertz, P. Gori Giorgi and S.M. Sutter for a careful reading of the manuscripts and helpful comments.

\appendix

\section{Derivation of the non-interacting entropy $S_0[\{n_p\}]$}
\label{app:deriventropy}
We start from the definition of the non-interacting free energy functional of equation \ref{eq:nonintfreeenergy} and write the corresponding Lagrangian,
\begin{align}
L_0^\beta[\gamma, \hat{\Gamma}_0, \{\epsilon_{pq}\}, \lambda] =&\quad \frac{1}{\beta} \Tr(\hat{\Gamma}_0 \log(\hat{\Gamma}_0))\\
&+ \sum_{pq} \epsilon_{pq} \left( \Tr(\hat{\Gamma} \hat{\gamma}_{pq}) - \gamma_{pq} \right)\\
&- \lambda \left(\Tr(\hat{\Gamma}_0) - 1 \right),
\end{align}
where $\epsilon_{pq}$ is the Lagrange multiplier for the 1-RDM and $\hat{\gamma}_{pq} = a^\dagger_p a_q$ in the general case and $\hat{\gamma}_{pq} = a^\dagger_{p\uparrow} a_{q\uparrow} + a^\dagger_{p\downarrow} a_{q\downarrow}$ for restricted singlet 1-RDMFT. The normalization constraint enforced through the Lagrange multiplier $\lambda$ is not necessary in the canonical ensemble as it can be simply absorbed in a constant shift in the diagonal of $\epsilon$, but we choose to retain it in the derivation. No positive semi-definiteness constraint on $\hat{\Gamma}_0$ is necessary due to the entropic term.\\
\\
We first take the derivative towards $\hat{\Gamma}_0$ and obtain,
\begin{equation}
\frac{\delta L_0^\beta}{\delta \hat{\Gamma}_0} = \frac{1}{\beta} \log(\hat{\Gamma}_0) + \frac{1}{\beta} + \sum_{pq} \epsilon_{pq} \hat{\gamma}_{pq} - \lambda = 0,
\end{equation}
from which it follows that,
\begin{equation}
\hat{\Gamma}_0 = e^{-\beta \sum_{pq} \epsilon_{pq} \hat{\gamma}_{pq} + \beta \lambda -1},
\end{equation}
which simplifies our Lagrangian considerably to
\begin{equation}
L_0^\beta[\gamma, \{\epsilon_{pq}\}, \lambda] = - \Tr(e^{-\beta \sum_{pq} \epsilon_{pq} \hat{\gamma}_{pq} + \beta \lambda  -1}) - \sum_{pq} \epsilon_{pq} \gamma_{pq} + \lambda.
\end{equation}
We now shift $\lambda \rightarrow \lambda + \frac{1}{\beta}$ to absorb the $-1$ in the exponent and obtain,
\begin{equation}
L_0^\beta[\gamma, \{\epsilon_{pq}\}, \lambda] = - \Tr(e^{-\beta \sum_{pq} \epsilon_{pq} \hat{\gamma}_{pq} + \beta \lambda}) - \sum_{pq} \epsilon_{pq} \gamma_{pq} + \lambda + \frac{1}{\beta},
\end{equation}
optimizing to $\lambda$ we obtain the normalization constraint,
\begin{equation}
e^{ \beta \lambda}\Tr(e^{-\beta \sum_{pq} \epsilon_{pq} \hat{\gamma}_{pq}}) = 1,
\end{equation}
which can be solved by setting,
\begin{equation}
e^{\beta \lambda} = \frac{1}{\Tr(e^{-\beta \sum_{pq} \epsilon_{pq} \hat{\gamma}_{pq}})}.
\end{equation}
We identify here our partition function,
\begin{equation}
Z = \Tr(e^{-\beta \sum_{pq} \epsilon_{pq} \hat{\gamma}_{pq}}) = e^{- \beta \lambda},
\end{equation}
and so our Lagrangian further reduces to
\begin{equation}
L_0^\beta[\gamma, \{\epsilon_{pq}\}] = - \frac{1}{\beta} \log(Z) - \sum_{pq} \epsilon_{pq} \gamma_{pq} + \frac{1}{\beta}.
\end{equation}
Optimizing towards $\epsilon_{pq}$ we obtain,
\begin{equation}
\frac{\Tr(a^\dagger_p a_q e^{- \beta \sum_{rs} \epsilon_{pq} a^\dagger_r a_s})}{Z} = \gamma_{pq}.
\end{equation}
Note that the matrix $\epsilon$ must be diagonal in the same basis as the 1-RDM (the NOs) to fulfill this constraint. Therefore, we switch to the NO basis and it becomes clear that our result only depends on the NOONs, and not on the NOs themselves. Our density matrix is given by $\hat{\Gamma} = \frac{e^{- \beta \sum_p \epsilon_p a^\dagger_p a_p}}{Z}$. The Lagrangian is then,
\begin{equation}
L_0^\beta[\{n_p\}, \{\epsilon_{p}\}] = - \frac{1}{\beta} \log(Z) - \sum_{p} \epsilon_{p} n_p + \frac{1}{\beta}.
\end{equation}
Our density matrix $\hat{\Gamma}_0$ is therefore diagonal in the basis of Slater Determinants constructed from the NOs. After finding the correct values of $\{\epsilon_p\}$ our non-interacting entropy can be found to be,
\begin{equation}
S_0[\{n_p\}] = \log(Z) + \beta \sum_{p} \epsilon_{p} n_p,
\end{equation}
and the corresponding non-interacting free energy is,
\begin{equation}
\FE_0^\beta[\{n_p\}] = - \frac{1}{\beta} \log(Z[\{\epsilon_p\}) - \sum_{p} \epsilon_{p} n_p.
\end{equation}
$\hat{\Gamma}_0$ is simply the density matrix of a non-interacting system with Hamiltonian $\hat{H}_0 = \sum_p \epsilon_p a^\dagger_p a_p$ and therefore we obtain the simple results in the grand canonical ensemble via the Fermi-Dirac distribution reported in equation \ref{eq:SGC} and \ref{eq:occGC}. The chemical potential in equation \ref{eq:occGC} is ``absorbed" into $\{\epsilon_p \}$.

\section{Connection to entropically regularized Multi-Marginal Optimal Transport}
\label{app:mmot}
The Multi-Marginal Optimal Transport problem is defined as,
\begin{equation}
C[\{n^M\}] = \min_{\Gamma \rightarrow \{n^1, n^2, \dots, n^N\}} \sum_{p_1p_2\dots p_N} \Gamma_{p_1p_2\dots p_N} \tilde{c}_{p_1p_2\dots p_N},
\end{equation}
where $n^M$ is the $M$th marginal, $\Gamma$ is referred to as the transport plan, while $\tilde{c}$ is the (transport) cost. The constraint $\Gamma \rightarrow \{n^1, n^2, \dots, n^N\}$ is given explicitly as,
\begin{equation}
\sum_{p_1p_2 \dots p_N / p_M} \Gamma_{p_1p_2\dots p_N} = n_{p_M}^M, \label{eq:marginals}
\end{equation}
where $\sum_{p_1p_2 \dots p_N / p_M}$ denotes the summation over all indices except $p_M$. The MMOT problem is a linear programming problem with a computational cost scaling \textit{in principle} exponentially with the number of marginals. However, with identical marginals and a pairwise symmetric cost $\tilde{c}_{p_1p_2\dots p_N} = \frac{1}{2}\sum_{i=1, j \neq i}^N c_{p_i p_j}$ it has been suggested that the problem may in fact be tractable computationally. \cite{FriVog-SJMA-18,FriSchVog-arxiv-21}\\
\\
One method of making the MMOT problem more tractable is to introduce an entropic regularization with inverse temperature $\beta$ in the following way,
\begin{align}
\FE^\beta[\{n^M\}] =&\,  \min_{\Gamma \rightarrow \{n^1, n^2, \dots, n^N\}} \\
& \Big( \sum_{p_1p_2\dots p_N} \Gamma_{p_1p_2\dots p_N} \tilde{c}_{p_1p_2\dots p_N} \nonumber\\
&+ \frac{1}{\beta} \sum_{p_1p_2\dots p_N} \Gamma_{p_1p_2\dots p_N} \log(\Gamma_{p_1p_2\dots p_N}) \Big) \nonumber.
\end{align}
An explicit expression for $\Gamma$ can then be found in terms of the Lagrange multipliers $\{\epsilon^M\}$ corresponding to the $N$ different marginals $\{n^M\}$,
\begin{equation}
\Gamma_{p_1p_2\dots p_N} = e^{- \beta \tilde{c}_{p_1p_2\dots p_N} - \beta \sum_{i=1}^N \epsilon_{p_i}}.
\end{equation}
The derivation is essentially identical to that of appendix \ref{app:deriventropy}, except for the fact that the normalization here is absorbed into the Lagrange multipliers $\{\epsilon^M\}$. The Sinkhorn algorithm is then obtained by inverting equation \ref{eq:marginals},
\begin{align}
\epsilon_{p_M}^{(n+1)} =&\, - \frac{1}{\beta} \log(n_p^M)\\
&+ \frac{1}{\beta} \log\left(\sum_{p_1p_2 \dots p_N / p_M} e^{- \beta \tilde{c}_{p_1p_2\dots p_N} - \beta \sum_{i\neq M} \epsilon_{p_i}^{i(n)}}\right).
\end{align}
As mentioned in the main text, to recover the correct result for non-interacting fermions, we take $\tilde{c}_{p_1p_2\dots p_N} = \sum_{i,j} c_{p_i p_j}$, with,
\begin{equation}
c_{pq} = \begin{cases} \infty & p = q\\
0 & p \neq q.
\end{cases}
\end{equation}
For non-interacting bosons the issue is slightly more complicated, because MMOT applied to particles in this manner assumes the particles to be distinguishable. If we take for example the case of $N=2$ with $c_{pq} = 0$, the state with both bosons in the same orbital $p=q$ is included once, while the state with $p \neq q$ is included twice. To adjust for the bosonic statistics, we modify the cost to effectively count the state with $p=q$ twice. The overall scaling factor is absorbed into normalization. The pairwise cost is then $c_{pq} = - \delta_{pq} \frac{\log(2)}{\beta}$, which provides the states for which $p=q$ with an additional factor of 2.\\
\\
However, for $N=3$ the pairwise cost does not give the correct result (the states with $p=q=r$ obtain a factor of $8$ instead of the desired $3! = 6$) and we must include an additional diagonal three-body cost $c_{pqr} = -\delta_{pq} \delta_{qr} \frac{\log(3/4)}{\beta}$. For $N=4$ one needs to introduce an additional diagonal four-body cost, and so on. In general an $N$-body cost $c_{p_1 p_2 \dots p_N} = - \delta_{p_1 p_2 \dots p_N} \frac{1}{\beta} \log(C_N)$ needs to be added to the cost for $N-1$ particles to reproduce the bosonic problem. In general we find $C_N = \frac{N!}{\prod_{M=1}^{N-1} C_M^{\binom{N}{M}}}$, where we take $C_1=1$. Then for example one finds $C_4 = \frac{32}{27}$, $C_5 = \frac{3645}{4096}$, $C_6 = \frac{67108864}{61509375}$, etc.\\
\\
Note also that the explicit presence of $\beta$ indicates that this strategy only works in the entropically-regularized case. As $\beta \rightarrow \infty$ the proposed costs become ill-defined. A more natural approach to identical particles is to reformulate the MMOT problem with a constrained summation, in the case of bosons,
\begin{equation}
C[\{n_p\}] =  \min_{\Gamma \rightarrow n} \sum_{p_1\leq p_2 \leq \dots \leq p_N} \Gamma_{p_1p_2\dots p_N} \tilde{c}_{p_1p_2\dots p_N},
\end{equation}
where the constraint is given explicitly by,
\begin{equation}
\sum_{M=1}^N \sum_{p_1 \leq p_2 \leq \dots \leq p_N / p_M = q} \Gamma_{p_1p_2\dots q \dots p_N} = n_{q},
\end{equation}
where with $\sum_{p_1 \leq p_2 \leq \dots \leq p_N / p_M = q}$ we mean that we set $p_M$ to equal $q$ in the summation, but still the inequalities on the indices must be respected. For example, for $N=2$ we would have,
\begin{equation}
\sum_{p_2 \geq q} \Gamma_{q p_2} + \sum_{p_1 \leq q} \Gamma_{p_1 q} = n_q.
\end{equation}
For fermions instead we need strict inequalities,
\begin{equation}
C[\{n_p\}] =  \min_{\Gamma \rightarrow n}\sum_{p_1 < p_2 < \dots < p_N} \Gamma_{p_1p_2\dots p_N} \tilde{c}_{p_1p_2\dots p_N},
\end{equation}
and so the density constraint is
\begin{equation}
\sum_{M=1}^N \sum_{p_1 < p_2 < \dots < p_N / p_M = q} \Gamma_{p_1p_2\dots q \dots p_N} = n_{q}.
\end{equation}
MMOT written in this manner and with entropic regularization is identical to the approach taken in the main text. In implementations of entropically-regularized Optimal Transport the term $\FE_N$ in equation \ref{eq:Sinkhorn} resulting from normalization is not included. Instead different marginals, and therefore different potentials (in this case: orbital energies) are assumed, which at every iteration adjust to give the correct normalization. In the case of identical marginals, we find it beneficial to include the normalization explicitly, to avoid having to fix the normalization after every iteration.

\section{Expressions for fermionic $W_0[\gamma]$ in different wavefunction subspaces}
\label{app:interactionenergy}
We start from equation \ref{eq:interactionenergy}, and we restrict ourselves here to singlet states, where $n_p^\uparrow = n_p^\downarrow = \frac{\tilde{n}_p}{2}$. Note that because of the anti-symmetry of $\langle pq||pq \rangle$ we can work directly with $\langle \hat{n}_p \hat{n}_q \rangle $ and neglect the $\delta_{pq}$ term. For the grand canonical ensemble ($\mathcal{F}$) we obtain the Restricted Hartree-Fock (RHF) expression of the interaction energy,
\begin{align}
W_{0, \mathcal{F}}[\gamma] =&\, \frac{1}{2} \sum_{pq} n_p n_q \langle pq || pq \rangle\\
&= \frac{1}{2} \sum_{pq} n_p^\uparrow n_q^\uparrow \langle pq || pq \rangle + \sum_{pq} n_p^\uparrow n_q^\downarrow \langle pq | pq \rangle \nonumber \\
&+ \frac{1}{2} \sum_{pq} n_p^\downarrow n_q^\downarrow \langle pq || pq \rangle\\
=&\, \frac{1}{4 }\sum_{pq} \tilde{n}_p \tilde{n}_q \langle pq | pq \rangle + \frac{1}{4} \sum_{pq} \tilde{n}_p \tilde{n}_q \langle pq || pq \rangle \nonumber \\
=&\, \frac{1}{2} \sum_{pq} \tilde{n}_p \tilde{n}_q \langle pq | pq \rangle - \frac{1}{4} \sum_{pq} \tilde{n}_p \tilde{n}_q \langle pq | qp \rangle,
\end{align}
which for RHF (all occupied $\tilde{n}_i = 2$) indeed reduces to
\begin{equation}
\sum_{ij} \big(2 \langle ij | ij \rangle - \langle ij | ji\rangle \big).
\end{equation}
For the canonical ensemble without spin restrictions ($\mathcal{H}_N$) we obtain,
\begin{align}
W_{0, \mathcal{H}_N}[\gamma] =&\, \frac{1}{2} \sum_{pq} \langle n_p n_q \rangle \langle pq || pq \rangle\\
=&\, \frac{1}{2} \sum_{pq} \langle n_{p\uparrow} n_{q\uparrow} \rangle \langle pq || pq \rangle + \sum_{pq} \langle n_{p\uparrow} n_{q\downarrow} \rangle \langle pq | pq \rangle \nonumber \\
&+ \frac{1}{2} \sum_{pq} \langle n_{p\downarrow} n_{q\downarrow} \rangle \langle pq || pq \rangle,
\end{align}
and no further simplification is possible. Restricting the ensemble only to the $\hat{S}_z = 0$ sector ($\mathcal{S}_z$) we obtain,
\begin{align}
W_{0, \mathcal{S}_z}[\gamma] =&\, \frac{1}{2} \sum_{pq} \langle n_p n_q \rangle \langle pq || pq \rangle\\
=&\, \frac{1}{2} \sum_{pq} \langle n_{p\uparrow} n_{q\uparrow} \rangle \langle pq || pq \rangle + \sum_{pq} n_{p\uparrow} n_{q\downarrow} \langle pq | pq \rangle \nonumber\\
&+ \frac{1}{2} \sum_{pq} \langle n_{p\downarrow} n_{q\downarrow} \rangle \langle pq || pq \rangle.
\end{align}
For the $N=2$ singlet the first and third term are zero, because we have $N_\uparrow = N_\downarrow = 1$ and so we are left with only,
\begin{equation}
W_{0, \mathcal{S}_z}[\gamma] = \sum_{pq} n_p^\uparrow n_q^\downarrow \langle pq | pq \rangle = \frac{1}{4}\sum_{pq} \tilde{n}_p \tilde{n_q} \langle pq | pq \rangle.
\end{equation}
Restricting ourselves to the $\hat{S}^2 =0$ sector ($\mathcal{S}$) we obtain,
\begin{align}
\langle \hat{\tilde{n}}_p \hat{\tilde{n}}_q \rangle &= \langle(\hat{n}_{p \uparrow} + \hat{n}_{p \downarrow})(\hat{n}_{q \uparrow} + \hat{n}_{q \downarrow}) \rangle\\
&=  \langle\hat{n}_{p \uparrow} \hat{n}_{q \uparrow}\rangle + \langle\hat{n}_{p \uparrow} \hat{n}_{q \downarrow}\rangle + \langle\hat{n}_{p \downarrow} \hat{n}_{q \uparrow}\rangle + \langle\hat{n}_{p \downarrow} \hat{n}_{q \downarrow}\rangle \nonumber\\
&= 2 \langle\hat{n}_{p \uparrow} \hat{n}_{q \uparrow}\rangle + 2 \langle\hat{n}_{p \uparrow} \hat{n}_{q \downarrow}\rangle.
\end{align}

To compute $W_0[\gamma]$ for $\mathcal{S}$, we use the fact that our two-body interaction in terms of singlet excitation operators only, we can obtain directly,
\begin{equation}
W_{0, \mathcal{S}}[\gamma] = \frac{1}{2} \sum_{pq} \left(\langle \hat{\tilde{n}}_p \hat{\tilde{n}}_q \rangle - \delta_{pq} \tilde{n}_p\right) \left( \langle pq | pq \rangle - \frac{1}{2}\langle pq | q p \rangle \right).
\end{equation}

\end{document}